\documentclass[aps,pra,reprint,superscriptaddress,showpacs]{revtex4-1}
\usepackage{graphicx}
\usepackage{bm}
\usepackage{amsmath}
\usepackage{bbold}

\usepackage[american]{babel}

\newcommand{\bra}[1]{\langle #1 |}
\newcommand{\ket}[1]{| #1 \rangle}

\begin{document}

\title{On the effective Dirac equation for ultracold atoms in optical lattices: \\
role of the localization properties of the Wannier functions}

\author{Xabier Lopez-Gonzalez}
\affiliation{\mbox{Depto. de F\'isica Te\'orica e Hist. de la Ciencia, Universidad del Pais Vasco UPV/EHU, 48080 Bilbao, Spain}}

\author{Jacopo Sisti}
\author{Giulio Pettini}
\affiliation{Dipartimento di Fisica e Astronomia, Universit\`a di Firenze,
and INFN, 50019 Sesto Fiorentino, Italy}

\author{Michele Modugno}
\affiliation{\mbox{Depto. de F\'isica Te\'orica e Hist. de la Ciencia, Universidad del Pais Vasco UPV/EHU, 48080 Bilbao, Spain}}
\affiliation{IKERBASQUE, Basque Foundation for Science, 48011 Bilbao, Spain}

\date{\today}

\begin{abstract}
We review the derivation of the effective Dirac equation for ultracold atoms in one-dimensional bichromatic optical lattices, following the proposal by [Witthaut \textit{et al.} Phys. Rev. A \textbf{84}, 033601 (2011)]. We discuss how such a derivation -- based on a suitable \textit{rotation} of the Bloch basis and on a 
\textit{coarse graining}  approximation -- is affected by the choice of the Wannier functions entering the coarsening procedure. We show that in general the Wannier functions obtained by rotating the maximally localized Wannier functions for the original Bloch bands can be sufficiently localized for justifying the coarse graining approximation. We also comment on the relation between the rotation needed to achieve the Dirac form and the standard Foldy-Wouthuysen transformation. Our results provide a solid ground for the interpretation of the experimental results by [Salger \textit{et al.} Phys. Rev. Lett. \textbf{107}, 240401 (2011)] in terms of an effective Dirac dynamics.
\end{abstract}

\pacs{67.85.Hj, 03.75.Lm, 03.65.Pm}
\maketitle

\section{Introduction}

The analog of Klein tunneling -- the penetration of relativistic-like particles through a potential barrier -- has been recently observed in a proof-of-principle experiment with ultracold atoms in a one-dimensional optical lattice \cite{salger}. This experiment follows a theoretical proposal by Witthaut \textit{et al.} \cite{witthaut} for simulating the $1+1$ Dirac equation 
in bichromatic optical lattices in the presence of a Dirac point, that is when the energy dispersion for a set of two Bloch bands takes the relativistic form $E_{\pm}(q)=\pm\sqrt{m^{2} c^{4} + c^{2}q^{2}}$. In fact, in this case it is possible to transform the original Schr\"odinger equation into a Dirac equation, by means of a suitable mixing (\textit{rotation}) of the two bands and of a \textit{coarse graining} procedure via a projection over a basis of Wannier functions \cite{witthaut}. 

A crucial point that guarantees the validity of  this reduction is the existence of a set of Wannier functions sufficiently localized within each lattice cell, in the \textit{rotated} basis \cite{note1}. In order to clarify this point, in this paper we present a detailed derivation of the Dirac effective equation, that allows to highlight the role played by the Wannier functions (that are not uniquely defined, owing to the arbitrariness of the phase of the Bloch functions \cite{marzari,modugno}), analyzing the specific cases discussed in Refs. \cite{salger,witthaut}. We show that even the Wannier functions obtained simply by rotating the maximally localized Wannier functions (MLWFs) for the original bands can be a reasonable choice. Our results provide a solid justification of  the good agreement between the experimental results of Salger \textit{et al.} \cite{salger} (or the numerical results of Ref. \cite{witthaut}) with the effective Dirac equation proposed by Witthaut \textit{et al.} \cite{witthaut}.

The article is organized as follows. In Sect. \ref{sec:effectivedirac} we discuss the derivation of the Dirac effective equation and the role played by the Wannier functions in the rotated basis. Then, in Sect. \ref{sec:fw} we discuss the relation between the rotation of the Bloch bands and the Foldy-Wouthuysen transformation for the Dirac equation.
In Sect. \ref{sec:mlwfs} we briefly review the concept of maximally localized Wannier functions, and discuss its relevance to the original and rotated Bloch basis. In Sect. \ref{sec:results} we explicitly compute the MLWFs for the original Bloch bands, and discuss how the rotation affects their localization properties.
There we consider explicitly both cases of the theoretical proposal by Witthaut \textit{et al.} \cite{witthaut} and of the experimental realization by Salger \textit{et al.} \cite{salger}. 
The implications for the sub-leading term in the expansion of the ``slowly-varying'' potential describing the potential barrier are examined in Sect. \ref{sec:potential}.  Concluding remarks are drawn in Sect. \ref{sec:conclusions}. 

\section{Effective Dirac dynamics}
\label{sec:effectivedirac}

Let us start from the single particle Schr\"odinger
equation  in the presence of a periodic potential $V_L({x})$ (of period $d$) and 
a slowly varying external potential $V({x})$
\begin{equation}
i\hbar\partial_t\Psi({x},t)=\left[
  H_L({x})+V({x})\right]
\Psi({x},t),
 \label{eq:schrod}
\end{equation}
where $H_L=-(\hbar^2/2M)\nabla^2+V_L$ is the unperturbed lattice Hamiltonian, whose eigenvectors 
are Bloch functions
$\psi_n({k},{x})={\rm
e}^{i{k}{x}}u_n({k},{x}) \equiv
\langle{x}|n,{k}\rangle$.
Then, Eq. (\ref{eq:schrod}) can be mapped to quasimomentum space as
 (see e.g. \cite{callaway,morandi})
\begin{align}
i\hbar\partial_t\varphi_n({k},t)&=E_n({k})\varphi_n({k},t)
\nonumber\\
&\qquad+
\sum_{n'}\int_{k'}\langle
n,{k}|V|n',{k}'\rangle\varphi_{n'}({k}',t)
\end{align}
where $\varphi_n({k},t)$ represent the expansion coefficients of a generic wave-packet $\Psi({x},t)$
on the Bloch basis, namely $\Psi({x},t)=
\sum_n\int_{{k}}\varphi_n({k},t)\psi_n({k},{x})$, and ${k}$ runs over the first Brillouin zone (the dependence on $t$ will be omitted in the following). 
The above equation can be written in vectorial form as 
\begin{equation}
i\hbar\partial_t\underline{\varphi}({k})=H_{L}({k})\underline{\varphi}({k})+
\int_{k'}\tilde{V}({k},{k}')\underline{\varphi}({k}'),
\label{eq:vec}
\end{equation}
with $H_{L}({k})=E_{n}({k})\delta_{nn'}$, $\tilde{V}({k},{k}')=
\langle n,{k}|V|n',{k}'\rangle$.

Let us now consider a subset of two bands, and assume that around $k=0$ the dispersion relation can be approximated as $E_{\pm}(k)=\pm\sqrt{m^{2} c^{4} + c^{2}(\hbar k)^{2}}$ (modulo an irrelevant constant), as considered in Ref. \cite{witthaut}. Then, in order to put the above expression in the form of a Dirac equation, it is convenient to make use of a $SO(2)$ rotation $R(\theta(k))$  \cite{witthaut}, with
\begin{equation}
R(\theta(k))= \begin{pmatrix} \cos\theta(k) & -\sin\theta(k) \\ \sin\theta(k) & \cos\theta(k) \end{pmatrix}
\label{eq:mixing}
\end{equation}
and
\begin{equation}
\tan\theta(k)=-\frac{m c^{2}}{c\hbar k + \sqrt{m^{2} c^{4} + c^{2}(\hbar k)^{2}}}.
\label{eq:theta}
\end{equation}
We notice that this rotation is related to an inverse Foldy-Wouthuysen transformation \cite{fw}; we shall come back to this point later on.
Then, Eq. (\ref{eq:vec}) can be written as
\begin{eqnarray}
i\hbar\partial_t \underline{\varphi}'({k})&=&
H_{L}'({k})\underline{\varphi}'({k})
\nonumber\\
&&
+\int_{k'}R({k})\tilde{V}({k},{k}') R^{T}({k}')\underline{\varphi}'({k}')
\label{eq:rotated}
\end{eqnarray}
with $\underline{\varphi}'=R\underline{\varphi}$, and 
\begin{equation}
H_{L}'({k})=R({k})H_{L}({k})R^{T}({k})
= \begin{pmatrix} c\hbar k & -m c^{2} \\ -m c^{2} & - c\hbar k \end{pmatrix}.
\label{eq:rotatedH}
\end{equation}

Equation (\ref{eq:rotated}) can be transformed back in coordinate space 
by projection on a basis of Wannier functions, as discussed in the following. We recall that the Wannier functions are obtained from the Bloch functions as \cite{callaway}
\begin{equation}
w_{n}({x}\!-\!{R}_i)=\sqrt{\frac{d}{2\pi}}
\int_k{\rm e}^{-i{k}{R}_i}\psi_n({k},{x}),
\label{eq:blochwannier}
\end{equation}
and that they are not uniquely defined due to the arbitrariness of the Bloch functions' phase (that, in general,  depends on $k$).
A generic wave packet $\Psi({x})$ can be expanded as $\Psi({x})=\sum_{n,i}
\chi_n({R}_i)w_{n}({x}\!-\!{R}_i)$, where the amplitudes $\chi_n({R}_i)$ can be
obtained from the Bloch coefficients by a
simple Fourier transform
\begin{equation}\label{eq:def_env_f}
\chi_n({R}_i)=\sqrt{\frac{d}{2\pi}}
\int_k\varphi_n({k}){\rm e}^{i{k}{R}_i}.
\end{equation}
The same relation holds in the rotated basis \cite{modugno}. 

When the Wannier functions (in the present case, those in the \textit{rotated} basis) are sufficiently localized in each cell, 
the rotated amplitudes $\chi'_n({R}_i)$ play the role of envelope
functions (associated to the site ${R}_i$, not just to the state $|w'_{n}(R_{i})\rangle$), corresponding to a \textit{corse graining} on the scale of a single cell \cite{adams,morandi}. 
Then, following the \textit{coarse graining} approximation,
 the coefficients 
$\underline{\chi}'({R}_i)$ can be supposed to be differentiable functions of ${R}_i$, the latter being considered as a continuous variable (eventually, $R_{i}\to x$). This holds when $\underline{\chi}'({R}_i)$ is slowly varying on the scale of the lattice period (``smooth'' wave packet). Under this approximation, and thanks to the properties of the Fourier transform \cite{adams,morandi}, 
the Hamiltonian $H_{L}'$ in coordinate space can be obtained by the replacement 
${k}\to -i{\nabla}_{{R}_i}$, so that Eq. (\ref{eq:rotated}) can be mapped in coordinate space as
\begin{eqnarray}
&&i\hbar\partial_t \underline{\chi}'({R}_i)=H_{L}'(-i{\nabla}_{{R}_i})\underline{\chi}'({R}_i)
\\&&+
\sum_{j}\int_{k}\int_{k'}e^{i{k}\cdot{R}_i}R({k})\tilde{V}({k},{k}')R^{T}({k}')e^{-i{k}'\cdot{R}_j}
\nonumber\underline{\chi}'({R}_j).
\label{eq:chiprime}
\end{eqnarray}
In addition, it is easy to show that
\begin{eqnarray}
&&\left.\int_{k}\int_{k'}R({k})\tilde{V}({k},{k}')R^{T}({k}')e^{i{k}\cdot{R}_i}e^{-i{k}'\cdot{R}_j}\right|_{nn'}
\nonumber\\
&&
=\int_{x}{w'_{n}}^{*}({x}-{R}_{i})V({x})w'_{n'}({x}-{R}_{j})\equiv \langle V\rangle^{ij}_{nn'},
\label{eq:10}
\end{eqnarray}
yielding
\begin{equation}
i\hbar\partial_t \underline{\chi}'({R}_i)=H_{L}'(-i{\nabla})\underline{\chi}'({R}_i)+
\sum_{j}\langle V\rangle^{ij}\underline{\chi}'({R}_j).
\label{eq:fulleq}
\end{equation}

When the Wannier functions are well localized inside each lattice cell, and the potential $V(x)$ is slowly varying on that scale, we can write 
\begin{equation}
\langle V\rangle^{ij}_{nn'}
\approx V({R}_{i})\delta_{nn'}\delta_{ij},
\label{eq:V-approx}
\end{equation}
so that we finally arrive at
\begin{equation}
i\hbar\partial_t \underline{\chi}'(x)=\left[ H_{L}'(-i{\nabla})+V(x)\right]
\underline{\chi}'(x).
\end{equation}
Then, the application of the $U(2)$ transformation \cite{witthaut}
\begin{equation}
U=\frac{1}{\sqrt{2}}\begin{pmatrix} 1 & -1 \\ 1 & 1 \end{pmatrix}
\label{eq:U}
\end{equation}
 yields ($\underline{\psi}\equiv U\underline{\chi}'$)
\begin{equation}
i\hbar\partial_t \underline{\psi}(x)= \begin{pmatrix} V(x) + m c^{2} & c \hat{p} \\ c \hat{p} & V(x) - m c^{2} \end{pmatrix}\underline{\psi}(x)
\label{eq:simildirac}
\end{equation}
that corresponds to the Dirac equation in $1+1$ dimensions, in the presence of a scalar potential $V(x)$. This is the same equation as obtained in \cite{witthaut}. We remark that in principle the transformation (\ref{eq:U}) could be applied before the coarse graining. This would affect the local behavior of the Wannier functions and therefore the subleading terms of the potential in Eq. (\ref{eq:fulleq}), but not the leading diagonal term $V(x)\mathbb{1}_{2\times2}$.

Summarizing, the present derivation shows that the mapping onto a Dirac equation is justified when there exists a set of sufficiently localized Wannier functions, in the \textit{rotated} basis \cite{note1}. Though they do not appear explicitly in the final expression (\ref{eq:simildirac}), they are needed to warrant both the coarse graining procedure and the expansion (\ref{eq:V-approx}) of the slowly varying potential \cite{note2}. The existence of Wannier functions with such properties will be discussed in the Sections \ref{sec:mlwfs} and \ref{sec:results}. 

\section{The Foldy-Wouthuysen transformation}
\label{sec:fw}

As anticipated in the previous section, the rotation (\ref{eq:mixing}) with the angle given by Eq. (\ref{eq:theta}) (also combined with the constant transformation $U$ in Eq. (\ref{eq:U})), corresponds to the inverse free-particle Foldy-Wouthuysen (FW) transformation in the momentum representation (that is, acting on the eigenstates of the Dirac equation) \cite{fw}.
We recall that the FW transformation is used to put the canonical \textit{free} Dirac equation into a convenient diagonal form, $H_{D}=\textrm{diag}(\sqrt{m^{2} c^{4} + c^{2}\hat{p}^{2}},-\sqrt{m^{2} c^{4} + c^{2}\hat{p}^{2}})$. This explains why here it is just the inverse FW transformation that allows to cast the original diagonal $2\times2$ Bloch Hamiltonian into the Dirac form. 

Owing to the above point, we notice that when the potential $V(x)$ is vanishing one  could even apply the coarse graining procedure directly in the original Bloch basis -- without rotation -- then claiming the equivalence with the Dirac equation via the inverse FW cited above (though, for practical purposes it might still be convenient to work with the canonical form, that is linear in the momentum operator $\hat{p}$.).  
However, the presence of the scalar potential $V(x)$ dramatically changes the situation. In fact, though one has still to use the same inverse free FW transformation to reshape the original Bloch Hamiltonian in quasimomentum space, the corresponding transformation after the coarse graining would not lead to the Dirac equation, as the potential $V(x)$ does not commute with the momentum operator $\hat{p}$. 
In this case the exact FW transformation is not known. Usually, it is customary to perform an expansion in $1/m$, leading to the well known spin-orbit and \textit{Zitterbewegung} terms of the non relativistic limit \cite{fw}. 
However, such an expansion is not useful in the present case, as here one is not interested in a the non-relativistic limit, but rather the opposite (that is, simulating relativistic effects close to $m=0$). Indeed, the direct transformation that leads to Eq. (\ref{eq:simildirac}) preserves all the relativistic contributions.

\section{Maximally localized Wannier functions}
\label{sec:mlwfs}

Among all possible choices, there exists a special class of Wannier functions, the so-called maximally localized Wannier functions (MLWFs) introduced by Marzari and Vanderbilt \cite{marzari}. These functions,  obtained by means of a suitable unitary gauge transformation of the Bloch eigenfunctions, are defined as those with the minimal spread, and can be constructed for both single or composite bands. Their application to bichromatic optical lattices has been recently discussed in \cite{modugno}. 

Let us consider explicitly the case of two almost degenerate bands, that is relevant to the present discussion.
The single band MLWFs are obtained via a diagonal unitary transformation of the form $\textrm{diag}(e^{i\phi_{1}(k)},e^{i\phi_{2}(k)})$, and correspond to the exponentially decaying Wannier functions discussed by Kohn \cite{kohn}. In general, it is convenient to define also a set of \textit{generalized} MLWFs for composite bands, by means of  a suitable gauge transformation obtained by parametrizing the most general $2\times2$ unitary  matrix \cite{modugno}. 
In the present case, the situation is complicated by the presence of the constraint (\ref{eq:theta}) that fixes the mixing angle $\theta(k)$. In this case, the only freedom left is in the choice of the phases of the original Bloch functions (before the rotation). Indeed, it is easy to prove that any other 
choice would spoil the form of the Hamiltonian (\ref{eq:rotatedH}), introducing a different dependence on $k$.

Then, in order to define a set of \textit{generalized} MLWFs that satisfy the constraint (\ref{eq:theta}), one could proceed as follows.
Given an initial set of Bloch functions $u_{n}(k,x)$ (that we suppose to be smooth functions of $k$ \cite{modugno}), the full transformation that minimizes the Wannier functions by preserving the Dirac form is
\begin{equation}
U(k)=R(\theta(k))\times\textrm{diag}(e^{i\phi_{1}(k)},e^{i\phi_{2}(k)})
\label{eq:gauge}
\end{equation}
(in principle, one could also include the constant $U(2)$ transformation, see Sect \ref{sec:effectivedirac}).
Following Ref. \cite{modugno}, the gauge dependent part of the Wannier spread, $\Omega_{U}$, can be expressed in terms of the generalized Berry vector potentials $A_{nm}(k)=i({2\pi}/{d})\bra{u_{nk}}\partial_{k}\ket{u_{mk}}$ as 
\begin{equation}
\Omega_{U}=\sum_{n=1,2}\langle \left(A_{nn}(k)-\langle A_{nn}\rangle_{\cal{B}}\right)^{2}\rangle_{\cal{B}}+2\langle |A_{12}|^{2}\rangle_{\cal{B}}.
\end{equation}
Generally, the two contributions in the above expression can be minimized either simultaneously or independently, and in one dimension they can be made strictly vanishing, in the so-called \textit{parallel transport} gauge \cite{modugno}.
However, since we want to preserve the form of the Hamiltonian (\ref{eq:rotatedH}), in the present case we can only require $\Omega_{U}$ to be minimum under the transformation (\ref{eq:gauge}), namely for
\begin{equation}
A_{nm}\rightarrow{\tilde A}_{nm}=i\sum_{l}U^{*}_{nl}{\partial_{k} U_{ml} }+\sum_{l,l'}U^{*}_{nl}U_{ml'}A_{ll'}.
\end{equation}
Notice that the diagonal gauge transformation in Eq. (\ref{eq:gauge}) only affects the diagonal term containing $A_{nn}$, leaving unchanged the off diagonal term $|A_{12}|$ \cite{modugno}. The latter is therefore fixed by the rotation $R(\theta)$. As a matter of fact, ${\tilde \Omega}_{U}$ results in a complicated integro-differential expression, whose solution is very tough, even numerically. Therefore, we shall adopt a different approach, as discussed in the following.

\section{The case of Refs. \cite{salger,witthaut}}
\label{sec:results}

As anticipated before, the coarse graining procedure is justified when the rotated Wannier functions are sufficiently localized on the scale of the lattice spacing, not necessarily those maximally localized. 
So, a sufficient condition is to start with the single band MLWFs for the original Bloch bands as considered by Witthaut \textit{et al.} \cite{note3}, and verify that the rotation (\ref{eq:mixing}) does not affect substantially their localization properties. In order to be quantitative, we shall use the normalized participation ratio
\begin{equation}
P=\left(d\int dx |w_{n}|^{4}\right)^{-1}
\end{equation}
as a measure of the extent of the Wannier functions $w_{n}(x)$, in units of the lattice period $d$.

Let us consider as examples the specific cases discussed in Refs. \cite{salger,witthaut}, with the periodic potential taking the form
\begin{equation}
V_{L}(x)=\frac{V_{1}}{2}\cos(2k_{L}x)+\frac{V_{2}}{2}\cos(4k_{L}x+\phi).
\end{equation}
As for the potential amplitudes, here  we consider two different sets, namely $V_{1}=5E_{R}$ and $V_{2}=1.6E_{R}$ \cite{salger} or $V_{2}=1.56E_{R}$ \cite{witthaut},  with $E_{R}=\hbar^{2}k_{L}^{2}/(2M)$ being the lattice recoil energy (whose actual value is irrelevant here). The two values of $V_{2}$ are very close but present subtle differences, as we shall discuss later on. 

The Bloch spectrum can be computed by a standard Fourier decomposition \cite{modugno}. Once the Bloch functions have been (numerically) obtained, one can compute the single band MLWFs by determining the phases $\phi_{1}(k)$ and $\phi_{2}(k)$  of the diagonal gauge transformation in Eq. (\ref{eq:gauge}), see e.g. Ref. \cite{modugno}, and then using Eq. (\ref{eq:blochwannier}). At the same time, one can also evaluate the \textit{rotated} MLWFs by using the full transformation (\ref{eq:gauge}), with the parameters $mc^{2}$ and $c$ obtained from a fit of the energy dispersion around $k=0$ \cite{witthaut}. Regarding this point, we have found that 
in the present regime of parameters, the Bloch bands cannot be exactly reproduced by the dispersion relation $E_{\pm}(k)=\pm\sqrt{m^{2} c^{4} + c^{2}\hbar^{2}k^{2}}$ (near $k=0$) as the independent fit of the two bands returns two different values for $c$ (the value of $mc^{2}$ is unambiguously fixed by the energy gap at $k=0$). However, in practice, by taking the average value of the effective velocity, $c\equiv (c_{+}+c_{-})/2$ we get a reasonable description of the exact Bloch bands. The actual values that we get for $mc^{2}$ and $c$ for the two cases in Refs. \cite{salger,witthaut} are reported in Tab. \ref{tab:mc} for different values of the phase (namely $\phi=0, 0.8\pi,\pi$).
\begin{table}[h!]
\begin{center}
\begin{tabular}{| c | c | c | c | c |}
\hline
\multicolumn{2}{|c|}{} & $\phi=0$ & $\phi=0.8\pi$ & $\phi=\pi$ \\ \hline
$V_{2}=1.56$ &$mc^{2}$ & 0.68 & 0.21 & $5\cdot10^{-4}$ \\ \hline
	& $c$ & 3.79 & 3.72 & 3.72 \\ \hline
$V_{2}=1.6$ & $mc^{2}$ & 0.69 & 0.21 & $8\cdot10^{-3}$ \\ \hline
	& $c$ & 3.79 & 3.72 & 3.72 \\ \hline
	\end{tabular}
\caption{Values of the parameters $mc^{2}$ and $c$ for different values of phase $\phi$ and $V_{2}=1.56E_{R},1.6E_{R}$ ($V_{1}=5E_{R}$). The values in the table are in lattice units (namely, energies in units of $E_{R}$, velocities in units of $E_{R}/(\hbar k_{L})$).}
\label{tab:mc}  
\end{center}
\end{table}

\begin{figure}[t!]
\centerline{\includegraphics[width=\columnwidth]{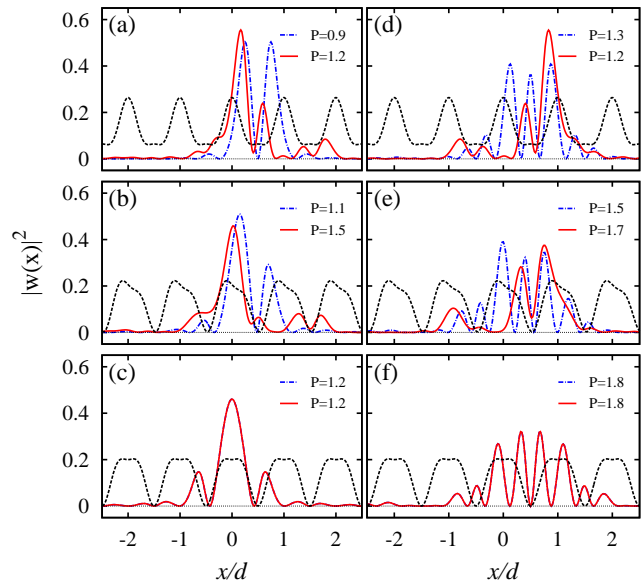}}
\caption{(Color online) Density plot of the Wannier functions, for the first and second excited bands (left and right, respectively) and $\phi=0,0.8\pi,\pi$ (from top to bottom). The MLWFs for the original Bloch bands are shown in blue (dotted-dashed line), those rotated in red (solid line). 
Here we use the set of parameters of Salger \textit{et al.} \cite{salger}: $V_{1}=5E_{R}$, $V_{2}=1.6E_{R}$. The numbers in the legend correspond to the values of the participation ratio $P$ (see text).}
\label{fig:mlwf-s}
\end{figure}
\begin{figure}[h!]
\centerline{\includegraphics[width=\columnwidth]{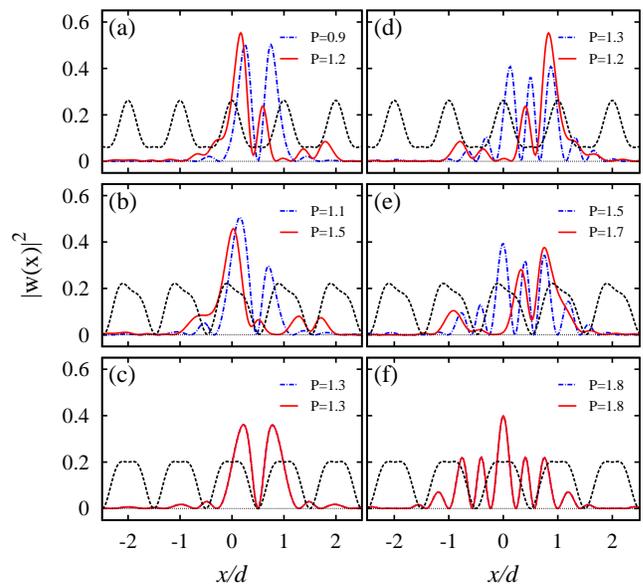}}
\caption{
(Color online) Density plot of the Wannier functions, for the set of parameters of Witthaut et al. \cite{witthaut}: $V_{1}=5E_{R}$, $V_{2}=1.56E_{R}$. See Fig. \ref{fig:mlwf-s} for comparison (and description).}
\label{fig:mlwf-w}
\end{figure}

The corresponding Wannier functions are shown in Figs. \ref{fig:mlwf-s} and \ref{fig:mlwf-w} (for the cases of Refs. \cite{salger,witthaut} respectively). In particular, there we show the single-band MLWFs (blue dotted-dashed lines), and the corresponding \textit{rotated} MLWFs defined above (red solid lines). 

A number of comments are in order. 
Firstly, it is obvious that in the ``relativistic'' \cite{salger} case $\phi=\pi$, the mass $m$ is almost vanishing, so that actually there is no rotation (see Eq. (\ref{eq:theta})), and the two sets of Wannier functions coincide (panels (c),(f)). 

Then, it is noteworthy that -- despite the similar values of the parameters -- the cases of Refs. \cite{salger,witthaut} present a different behavior, especially close to $\phi=\pi$ (panels (c),(f)). This is due to the fact that the two values  $V_{2}=1.56E_{R}$ and $V_{2}=1.6E_{R}$ lie on different sides with respect to the degeneracy point corresponding to the exact crossing of the two bands, that we numerically locate at $V_{2}\simeq1.5625 E_{R}$.
As a consequence, the $p$-like solution centered at the deepest minima of the cell and the $s$-like centered at the tiny minimum on top of the potential exchange their role when crossing the resonance (that is $E_{s}<E_{p}$ above the resonance, in Fig. \ref{fig:mlwf-s}, and vice versa below the resonance, in Fig. \ref{fig:mlwf-w}).

Finally, the most important remark regards the localization properties of the Wannier functions. Remarkably, Figs. \ref{fig:mlwf-s} and \ref{fig:mlwf-w} show that the rotation does not affect dramatically their localization properties, the \textit{rotated} Wannier functions having a behavior similar to the MLWFs for the original Bloch band. As a matter of fact, though the two sets of Wannier functions have a different ``microscopic'' structure, the corresponding values of the participation ration $P$ are not so different.  

In order to complete the analysis, one has to check how the approximation (\ref{eq:V-approx}) behaves under the rotation.
 This requires  a precise analysis of the sub-leading terms in Eq. (\ref{eq:fulleq}), that we shall discuss in the following section.

\section{The ``slowly varying'' potential}
\label{sec:potential}

Let us consider a generic \textit{slowly varying} potential (that used in Refs. \cite{salger,witthaut} takes the form $V(x)=V_{0}\exp[-2(x/x_{0})^{2}] -Fx$; the following treatment is valid in general). By performing a series expansion around $x=R_{j}$, the potential term in Eq. (\ref{eq:10}) can be written as
\begin{align}
\langle V\rangle^{jj'}_{nn'}&=\sum_{s}\frac{1}{s!}\left.\frac{\partial^{s}V}{\partial x^{s}}\right|_{R_{j}}\!\!\!\langle (x-R_{j})^{s}\rangle^{jj'}_{nn'}
\nonumber\\
&= V(R_{j}) + \left.\frac{\partial V}{\partial x}\right|_{R_{j}}\!\!\!\langle x-R_{j}\rangle^{jj'}_{nn'}+\dots
\nonumber\\
&= V(R_{j}) + \left.\frac{\partial V}{\partial x}\right|_{R_{j}}\!\!\!\left(\langle x\rangle^{jj'}_{nn'}-R_{j}\delta_{nn'}\delta_{jj'}\right)+\dots,
\nonumber
\end{align} 
so that the corrections to Eq. (\ref{eq:V-approx}) read 
\begin{align}
\delta V_{nn'}^{(\ell)}(R_{j})&\equiv\langle V({x})-V(R_{j})\rangle^{j,j+\ell}_{nn'}
\nonumber\\&=
\left.\frac{\partial V}{\partial x}\right|_{R_{j}}\!\!\!\left(\langle x\rangle^{(\ell)}_{nn'}-R_{j}\delta_{nn'}\delta_{\ell0}\right)+\dots
\end{align} 
with $\langle x\rangle^{(\ell)}_{nn'}$
being independent of $j$ owing to the invariance of the lattice under discrete translations.
Then, it is convenient to rescale the latter expression by the recoil energy $E_{R}$ and write it as
\begin{equation}
\frac{\delta V_{nn'}^{(\ell)}(R_{j})}{E_{R}}\approx\frac{d}{E_{R}}\left.\frac{\partial V}{\partial x}\right|_{R_{j}}\!\!\!\cdot\frac{1}{d}\left(\langle x\rangle^{(\ell)}_{nn'}-R_{j}\delta_{nn'}\delta_{\ell0}\right),
\end{equation}
where the term $(d/E_{R})(\partial V/\partial x)|_{R_{j}}$ represents the variation of the potential on the scale of the lattice spacing $d$, divided by the characteristic energy scale $E_{R}$ of the lattice. By hypothesis, this term is small under the assumption of a slowly varying potential. Then, in order to verify that the approximation (\ref{eq:V-approx}) is justified, one has to check that the remaining term
\begin{equation}
\Delta^{(\ell)}_{nn'}\equiv(\langle x\rangle^{(\ell)}_{nn'}-R_{j}\delta_{nn'}\delta_{\ell0})/d
\end{equation}
is sufficiently smaller than unity (note that $\Delta^{(\ell)}_{nn'}$ is actually independent of $j$ owing to the invariance of the lattice under discrete translations). In principle, one may expect this condition to be satisfied when the Wannier functions are sufficiently localized within each lattice cell. Indeed, we have  verified that  
$|\Delta^{(\ell)}_{nn'}|<0.5$ for $\ell=0,\pm1,\pm2$ for all the cases shown in Figs. \ref{fig:mlwf-s},\ref{fig:mlwf-w}.
Notice also that the actual value of the on-site diagonal term $\Delta^{(0)}_{nn}\equiv(\langle x\rangle^{(0)}_{nn}-R_{j})/d$ (the only one depending on $R_{j}$) is not univocally determined (though smaller than unity, anyway) due to the arbitrariness in choosing the origin of the unit cell.

\section{Conclusions}
\label{sec:conclusions}

We have revisited the derivation of the effective Dirac equation for non-interacting ultracold atoms in optical lattices, discussing in particular the role played by the localization properties of the Wannier functions entering the \textit{coarse graining} approximation. Though, remarkably, the choice of the Wannier functions does not appear explicitly in the final Dirac equation at leading order, the existence of a set of Wannier functions sufficiently localized within each lattice cell, is a crucial requirement.
We have shown that the Wannier functions must be calculated from the \textit{rotated} Bloch basis, and that a reasonable option can be obtained by rotating the MLWFs for the original Bloch bands.
The above results eventually justify the use of the \textit{coarse graining} approach, and account for the good agreement between the experimental results of Salger \textit{et al.} \cite{salger} (or the direct solution of the Schr\"odinger equation discussed in Ref. \cite{witthaut}) and the effective Dirac equation proposed by Witthaut \textit{et al.} \cite{witthaut}.

\acknowledgments
We acknowledge useful discussion with A. Barducci, A. Bergara, I. Egusquiza, G. Muga.
This work has been supported by the UPV/EHU under program UFI 11/55, the Spanish Ministry of Science and Innovation through Grant No. FIS2012-36673-C03-03, and the Basque Government through Grant No. IT-472-10.


\begin{thebibliography}{}
\bibitem{note1}
In Ref. \cite{witthaut}, only the properties of the (maximally localized) Wannier function of the original Bloch bands have been discussed.
\bibitem{note2}
The Wannier functions do not appear explicitly at the leading order as they are integrated out by the coarse graining procedure; however they would affect the next-to-leading corrections of the potential.
\end{thebibliography}
\end{document}